\documentclass{elsart}

\usepackage{harvard}

\usepackage{graphicx}

\usepackage{amssymb}

\def\url#1{{\ttfamily\def\/{/\discretionary{}{}{}}#1}}

\def\HI {H\kern0.1em{\sc i}}

\begin{document}

\begin{frontmatter}
\title{New VLBA Identifications of Compact Symmetric Objects}


\author{A. B. Peck\thanksref{abp}\thanksref{email}}, 
\author{G. B. Taylor\thanksref{email}}

\thanks[abp]{Present address: MPIfR, Auf dem H\"{u}gel 69, D-53121 Bonn, Germany}
\thanks[email]{E-mail: apeck@nrao.edu, gtaylor@nrao.edu}

\address{NRAO, P.O. Box O, Socorro, NM 87801}

\begin{abstract}

The class of radio sources known as Compact Symmetric Objects (CSOs)
is of particular interest in the study of the evolution of radio
galaxies.  CSOs are thought to be young (probably $\sim$10$^4$ years), and a
very high fraction of them exhibit \HI\ absorption toward the central
parsecs.  The \HI, which is thought to be part of a circumnuclear torus
of accreting gas, can be observed using the VLBA with high enough
angular resolution to map the velocity field of the gas.  This
velocity field provides new information on the accretion process in
the central engines of these young sources.
 
We have identified 9 new CSOs from radio continuum observations for
the VLBA Calibrator Survey, increasing the number of known CSOs by
almost 50\%.

\end{abstract}

\end{frontmatter}

\section{Introduction}
\label{intro}

Compact Symmetric Objects are distinguished from other classes of
compact radio sources by their striking degree of symmetry.  This is
due to the fact that CSOs tend to lie close to the plane of the sky,
and so relativistic beaming plays a very minor role in their observed
properties. They usually have well-defined lobes and edge-brightened
hotspots on either side of an active core, often exhibiting an "S"
shaped symmetry \cite{Tay96}.  CSOs are $<1$ kpc in size, and usually
have high intrinsic radio luminosities and low levels of polarization.
Their small physical size is thought to be attributable to the youth
of the sources, rather than to confinement by a dense medium
\cite{Rea96}.  It is likely that these sources will evolve into
Compact Steep Spectrum sources, and then into the large classical
doubles \cite{Rea96b}.  The advance speeds of the hotspots have been
measured for a few CSOs, using data taken over $\sim$10 years
\cite{Ows98}, and yield kinematic ages of 10$^3$ to 10$^4$ years.

One of the most important properties of CSOs for studies of galaxy
evolution is the very high rate of detection of \HI\ absorption lines \cite{Pec99a}.
In many cases, this \HI\ appears to be part of a circumnuclear
structure which is thought to be feeding and obscuring the central
engine \cite{Pec99c}.  For this reason, we have undertaken a survey to
identify new CSOs from the VLBA Calibrator Survey (VCS)
\citeaffixed{Pec97}{http://magnolia.nrao.edu/vlba\_calib/,}.

\begin{figure}
\begin{center}
\includegraphics*[width=13cm]{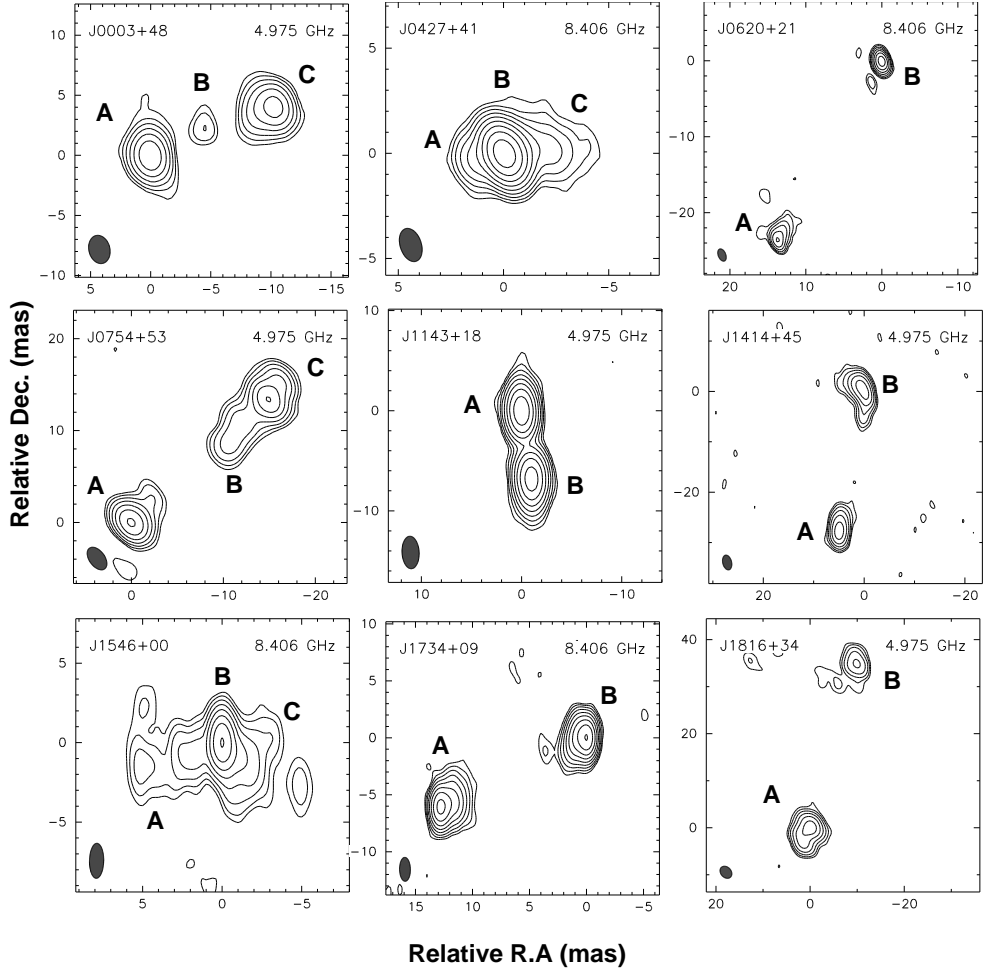}
\end{center}
\caption{Continuum images of the nine most promising CSO candidates.}
\label{peckb_fig1}
\end{figure}

\begin{figure}
\begin{center}
\includegraphics*[width=13cm]{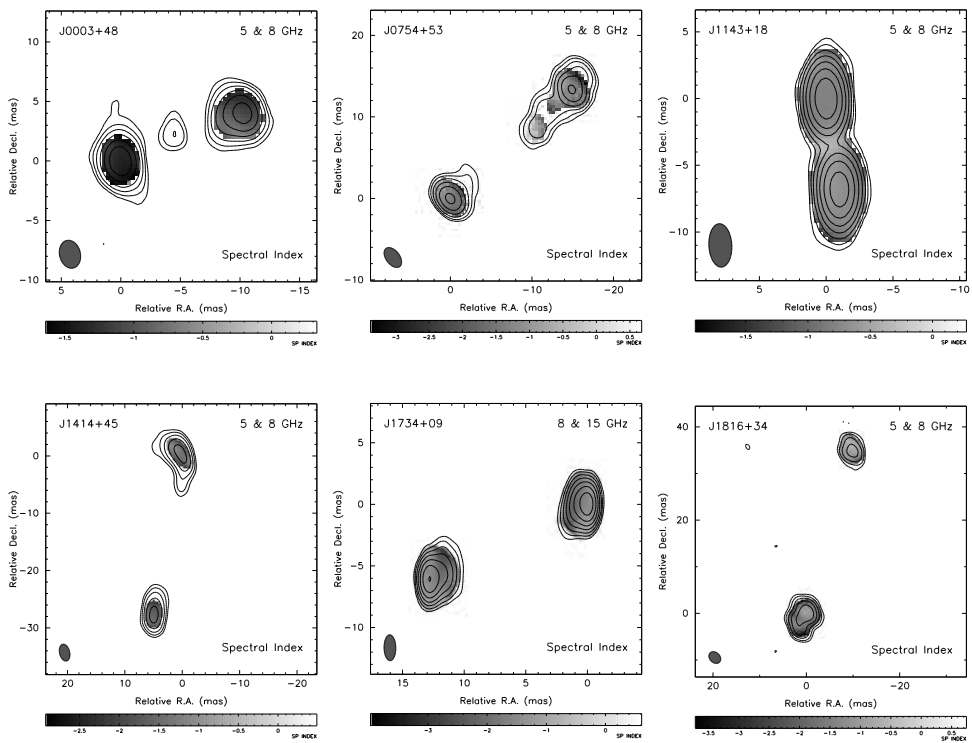}
\end{center}
\caption{Spectral Index distributions for six of the new CSOs.}
\label{peckb_fig2}
\end{figure}


\begin{table*}
\caption{Fluxes and Spectral Indices of CSOs}
\label{peck_tab1} 
\begin{center}
\begin{tabular}{ccccccc}
\hline 
Source &Comp. &$S_5$ &$S_{8.4}$ &$S_{15}$ &$\alpha^{5}_{8.4}$
 &$\alpha^{8.4}_{15}$  \\
(1) &(2) & (3)  &(4) &(5) &(6) &(7) \\
\hline 
J0003+4807 &A&  70.6 &36.9 &... &-1.24  &... \\
 &B     &   3.5&3.4 &... & -0.06&... \\
 &C & 66.0& 40.3& ...& -0.94&...  \\
J0427+4133 &A&... &32.7&22.0&... &-1.46 \\
 &B &... &588.9 &406.0 &... &-0.63 \\
 &C &... &28.7 &13.9 &... &-1.23  \\
J0620+2102 &A&478.4&77.2&...&... &... \\
 &B &400.0 &148.4 &...&... &... \\
J0754+5324 &A&76.1&29.4&... &-1.81&... \\
 &B  &26.7 &10.2 &... &-0.52 &...  \\
 &C  &79.6 &34.2 &... &-1.61 &...  \\
J1143+1834 &A&180.7&121.2&... &-0.76&... \\ 
&B &159.8 &112.7 &... &-0.67 & ... \\
J1414+4554 &A&76.1&34.2&... &-1.52&... \\
 &B &97.8 &41.9 & ...&-1.62 &... \\
J1546+0026 &A&...&94.2&60.7&... &-0.75  \\
 &B  &... &354.0 &242.9 & ...&-0.64 \\
 &C  &... &87.8 &71.7 & ...&-0.34 \\
J1734+0926 &A&...&162.4&66.7&... &-1.51 \\
 &B  &... &238.8 &97.2 &... &-1.53  \\
J1816+3457 &A&196.8&74.4&... &-1.85&...  \\
 &B &75.9 &27.7 &... &-1.92 &...  \\
\hline 
\end{tabular}
\end{center}

\end{table*}

\section{Results}
\label{results}
The sources in this study were selected from images of positive
declination sources in the VCS.  Negative dec. sources will be added
at a later date.  The CSO candidates were identified based on at least
one of the following criteria: a) double structure at 2 GHz, 8 GHz or
both, where ``double structure'' is considered to mean having two
distinct components with an intensity ratio $<10:1$; b) a strong
central component with possible extended structure on both sides at
one or both frequencies; c) possible edge-brightening of one or more
components.

Positive classification of CSOs requires multifrequency observations
in order to identify the expected steep spectrum hotspots and a flat
or inverted spectrum core.  The $\sim$40 selected sources were
observed with the VLBA at 2 frequencies in addition to the discovery
frequency.  Details of these follow-up observations are reported in
\citeasnoun{Pec99b}.

Images of the most promising CSO candidates are shown in Figure
\ref{peckb_fig1}.  The frequency of the image shown is indicated in
the upper right corner of each plot, and the beam is displayed in the
lower left.   Table \ref{peck_tab1} lists the flux densities of each
component at all frequencies at which follow-up observations were
made.  Columns 6 and 7 list the spectral indices of the components.
The spectral index distribution for sources in which such an image was
feasible are shown in Figure \ref{peckb_fig2}.

\section{Future Work}
\label{future}
VLBA imaging of \HI\ absorption in CSOs provides an unparalleled means
of determining the kinematics of neutral gas within 100 pc of an
active nucleus \cite{Pec99c,Tay99}.  VLBA imaging of free-free absorption in
CSOs is also possible, and allows us to probe the region of ionized
gas surrounding the central engine \cite{Pec99c}.  A complete sample
of such sources will yield unique information about accretion
processes and the fueling mechanism by which these young radio
galaxies might evolve into much larger FRII type sources.


\end{document}